\documentclass[aps]{revtex4}

\usepackage[english]{babel}
\usepackage[ansinew]{inputenc}
\usepackage{amssymb}                                                           % AMS math symbols.
\usepackage{amsmath}                                                           % AMS extensions.
\usepackage{graphicx}                                                          % Include figure files.
\usepackage{bm}                                                                % For bold math.

\newcommand{\ave}[1]{\langle{#1}\rangle}                                      % Statistical average.
\begin{document}

%%
%%---- Title of the paper --------------------------------------------------------------------------------
%%

\title{The Stochastic Gross-Pitaevskii Equation and some Applications}

%%
%%---- Authors and affiliations --------------------------------------------------------------------------
%%

\author{S.P. Cockburn}\email[E-mail: ]{s.p.cockburn@ncl.ac.uk}
\author{N.P. Proukakis}\email[E-mail: ]{nikolaos.proukakis@ncl.ac.uk}
\affiliation{School of Mathematics and Statistics, Newcastle University, Newcastle upon Tyne, NE1 7RU, United Kingdom}

\date{5 August 2008}

%%
%%---- Abstract ------------------------------------------------------------------------------------------
%%

\begin{abstract}
The stochastic Gross-Pitaevskii equation represents a versatile approach for studying the
dynamics of trapped degenerate ultracold Bose gases in the presence of large phase and density fluctuations. Following a brief review of the original formulation of Stoof, which highlights the benefits of this approach and its relation to alternative theories, we present selected applications for the dynamics of effectively one-dimensional systems, and briefly discuss the generalization to two-dimensional systems, highlighting certain potential pitfalls in their numerical implementations.
\end{abstract}

%%
%%---- PACS numbers --------------------------------------------------------------------------------------
%%

%\pacs{}
%% 03.75.Nt : Other Bose-Einstein condensation phenomena
%% 03.75.Lm : Tunneling, Josephson effect, Bose-Einstein condensates in periodic potentials, 
%%            solitons, vortices, and topological excitations
%% 05.10.Gg : Stochastic analysis methods
\maketitle

%%
%%---- Introduction --------------------------------------------------------------------------------------
%%
%\newpage
\section{Introduction}
The ultracold atomic samples now routinely produced in many laboratories worldwide, provide excellent means by which to investigate many-body quantum systems on a macroscopic scale. From a theoretical perspective, these systems are appealing as they support an array of rich dynamics, in many cases readily amenable to experimental observation. With this close relation between theory and experiment \cite{Stringari_Review,General_Review_2}, many such experimental features have been subject to analytical and numerical investigations: much attention has been paid to numerical analyses based upon the Gross-Pitaevskii equation (GPE) in particular \cite{Frantzeskakis_Book}, which has been shown to successfully describe condensate dynamics in the limit when the many-body wavefunction represents the condensate alone, thus corresponding to the low-temperature regime.

Most experiments on trapped ultracold gases feature, to a greater or lesser extent, a depletion of the condensate due to the effect of finite temperature. While in three dimensional systems, finite temperature effects play an increasingly important role as the critical temperature is approached, for low dimensional systems, thermal effects harbour greater importance still, due to the existence of enhanced phase fluctuations within systems under tight confinement in one or two directions 
\cite{Low_D_Experiment_1,Low_D_Experiment_2,Low_D_Experiment_3}. Moreover, the dynamics of the thermal component can play a significant role, for example in the case of an ultracold gas under strong external perturbations \cite{Jackson_Zaremba_1,Jackson_Zaremba_4,Excitations_ZNG,Jackson_LasPhys}.

A number of finite-temperature techniques have been developed to incorporate the effects of temperature into a description of trapped Bose gases, as recently summarised in \cite{MyReview,MyTutorial,NZ_OZ_tutorial,Yukalov_SB}, which feature a detailed comparison of the relative weaknesses and merits of such approaches. 
In brief, various theories have been put forward and applied to the experiments, as discussed below: In mean-field type approaches, the condensate is treated as a classical object, separated from the effect of the thermal cloud; various such perturbative approaches have been formulated to second order in the effective interaction strength 
\cite{ZNG,Bijlsma_Zaremba_Stoof,JILA_Kinetic_1,Proukakis_JPhysB}, with the common feature of obtaining a set of coupled dynamical equations for the condensate and the non-condensate, which interact both via mean fields and by particle exchange.
The condensate obeys an appropriately generalised GPE, which,
in the most commonly adapted version following from early work of Kirkpatrick and Dorfman \cite{Kirkpatrick_1,Kirkpatrick_2,Kirkpatrick_3}, is coupled to a quantum Boltzmann equation for the thermal cloud; this model has been extensively used by Zaremba, Nikuni, Griffin and co-workers, who introduced the terminology `ZNG' \cite{ZNG}.
Although such approaches work remarkably well in predicting a large range of experimentally-relevant phenomena, they do not fully describe regimes where fluctuations in the condensate phase are large, as is appropriate for example in low-dimensional systems
\cite{Low_D_Experiment_1,Low_D_Experiment_2,Low_D_Experiment_3}.
To incorporate such effects, generalised zero temperature \cite{Low_D_Petrov} and finite temperature \cite{Low_D_Stoof} modified methods were developed, to describe the static properties of such systems in an \emph{ab initio} manner. 

However, in order to go beyond mean field effects in a dynamical manner, various alternative approaches have been developed and implemented. In first instance, the time-dependent mean field approach mentioned above was reformulated in a somewhat more justified number-conserving manner \cite{Morgan_Gardiner}, following from early work \cite{Girardeau_Arnowitt,Gardiner_NC,Castin_Dum}; although this has led to good agreement with experiments in at least one case \cite{Excitations_Burnett_2003}, a full dynamical model of such theories is still lacking.

Classical field approaches are based upon the observation that \cite{Svistunov_3,Polish_Review}, as an equation for a classical field, the GPE should be able to describe the classical aspects of an ultracold trapped Bose gas. This includes all highly occupied modes, for which it is possible to approximate the field operators for the creation/annihilation of particles, by c-number amplitudes \cite{PGPE_Trap}. Consistency requires only the highly occupied modes to be propagated this way, which can be achieved by introducing a projector to the GPE \cite{PGPE}. Temperature effects can be incorporated through the initial conditions, which are typically set to some non-equilibrium distribution, and subsequently allowed to equilibrate at some temperature\cite{PGPE_Trap}, via the (projected-)GPE, under the restriction of fixed particle number and energy.

A variant of this technique is known as the Truncated Wigner method, originally developed in the context of quantum optics and first used for Bose gases in \cite{Steel_TWA}; a key element of this method is the approximate incorporation of quantum effects, 
%such as the zero temperature depletion of the condensate due to interactions, 
through the addition of a prescribed amount of quantum noise to the wavefunction initial conditions. 
The numerical procedure which results, is to propagate a set of stochastic matter fields, which serve to sample the Wigner distribution under this approximation, via an `appropriate' equation of motion for the  Wigner distribution. The terminology `truncated' Wigner arises from the fact that the propagation is carried out in an approximate manner, via the (projected) GPE.
A limitation of this approach is its inability to accurately describe thermal effects, although certain remedies of limited applicability have been proposed \cite{Castin_TWA}. In the latter two approaches, a temperature can only be assigned at the end of the simulations, with the thermalisation arising respectively due to ergodicity \cite{PGPE_Trap}, while simultaneously exhibiting an unphysical heating.

%The full equation of motion for the Wigner distribution is a partial differential equation involving first and third order derivative terms, however under the truncated Wigner approximation, only the first order terms are retained. The numerical procedure which results, is to propagate a set of stochastic matter fields, which serve to sample the Wigner distribution under this approximation, via the GPE. Again in doing so, the quantum field operator is assumed to be well approximated as a c-number field [Sinatra et al]. Thermal effects can also be incorporated through the addition of extra, appropriately distributed, noise to the initial wavefunction followed by propagation to equilibrium with the GPE and a suitably defined basis, as in [RobinScott+refs therein], for example. 

To gain better control over the temperature to which the system equilibrates, requires detailed consideration of the coupling between such modes, and the higher lying, thermal region in our system. Within such schemes, an appropriate generalisation has been formulated in \cite{PGPE_T}. Earlier, rather distinct, yet analogous in spirit,
means of achieving this was provided independently by Stoof 
\cite{Stoof_PRL,Stoof_JLTP,Stoof_Langevin,Stoof_Duine,Stoof_LesHouches}, and by Gardiner, Davis and co-authors \cite{QK_V,QK_VII,SGPE_I,SGPE_II,SGPE_III}, using very different theoretical approaches. 

This led to the generalisation of the GPE to a Langevin equation \cite{Stoof_Langevin}, often referred to as the Stochastic-Gross-Pitaevskii equation (SGPE) \cite{SGPE_II}, a terminology that we also adopt here; these two independent formulations have many formal similarities and are closely related, up to some subtle difference discussed in \cite{MyTutorial,SGPE_II,QK_VII}. The SGPE describes the stochastic evolution of the condensate plus highly occupied, low-lying thermal modes, under the influence of coherent and incoherent interactions with the particles occupying higher energy modes. 
%This is the theory on which we will focus here, the details of which will be expanded upon in the following section.

Given the diverse range of approaches currently available, our aim in this manuscript is to highlight the key elements of this latter approach, paying particular attention on some of its appealing features via somewhat idealised experimental scenarios. We start our discussion with a brief overview of one formulation of this approach, which has already attracted a relatively large literature 
\cite{Stoof_Langevin,Stoof_Duine,Low_D_PRA,SGPE_AtomLaser,Stoof_Vortex,Proukakis_AtomChip,Proukakis_Equilibrium}, with related work discussed in \cite{SGPE_II,SGPE_III,SGPE_IV,Davis_SGPE_New,Davis_SGPE_Nature}.

%%
%%---- Second paragraph -----------------------------------------------------------------------------------
%%

\section{Methodology}
\label{sec:method}

\subsection{The Stochastic Gross-Pitaevskii equation} 

The stochastic Gross-Pitaevskii equation is a non-equilibrium microscopic theory to describe the evolution of an ensemble of ultracold atoms in contact with a thermal cloud. Using the Keldysh non-equilibrium formalism, Stoof derived a Fokker-Planck equation for the the system in \cite{Stoof_JLTP}; such an equation is equivalent to the following Langevin equation \cite{Stoof_Langevin},
\begin{eqnarray} 
\label{Langevin}
i \hbar \frac{ \partial \Phi({\bf r};t) }{ \partial t} 
=  \Bigg[ - \frac {\hbar^{2} \nabla^{2} }{2m} + V_{\rm ext}({\bf r}) - \mu - iR({\bf r};t) 
%nonumber \\ & &
+ g |\Phi({\bf r};t)|^{2} \Bigg] \Phi({\bf r};t) + \eta({\bf r};t)\;,
\end{eqnarray}
where here $\Phi({\bf r};t)$ is the order parameter for the lowest, coherent system modes
 and the higher modes are assumed to obey their own coupled dynamics, being in general governed by a quantum Boltzmann equation as in \cite{ZNG,Bijlsma_Zaremba_Stoof,MyTutorial}.
%
%be populated according to equilibrium Bose-Einstein statistics. 
%
In addition to the usual Gross-Pitaevskii terms for the kinetic energy, trapping potential ($V_{\rm ext}({\bf r})$) and nonlinear interaction term (where $g=4\pi\hbar^{2}a_{3d}/m$, parameterised by the s-wave scattering length $a_{3d}$), this equation features two additional contributions: firstly, $R({\bf r};t)$ denotes a dissipative term which represents the coupling between the condensate and a thermal reservoir at a temperature $T$, and accounts for the transfer of particles between the high and low energy modes of the system; the contribution $\eta({\bf r};t)$ is a noise term which accounts for the random nature of incoherent collisions within the system and is somewhat analogous to the random particle jitter in Brownian-type motion (see, e.g.\ \cite{Stoof_Duine}). 
Finally, $\mu$ denotes an effective chemical potential.
The above noise contribution is characterized by Gaussian correlations of the form
\begin{equation}
\langle \eta^{*}({\bf r},t) \eta({\bf r'},t') \rangle = i (\hbar^{2}/2) \Sigma^{K}({\bf r},t) \delta (t-t') \delta ({\bf r}-{\bf r'}),
\end{equation}
where $\langle\ldots\rangle$ denotes averaging over different realisations of the noise (and it is generally understood that the results of simulations will undergo appropriate averaging, although single-runs can also offer qualitative results as discussed later). 
The amplitude of the noise is in general position and time dependent, and is governed by the so-called Keldysh self-energy $\Sigma^{K}({\bf r},t)$. For a thermal cloud which is sufficiently close to equilibrium, this takes the form \cite{Stoof_Langevin}
\begin{eqnarray}
\Sigma^{K}({\bf r},t) & = &
- i \left( \frac{4 \pi}{\hbar} \right) g^{2} \int \frac{ d {\bf p}_{2}}{(2 \pi \hbar)^{3}} 
\int \frac{ d {\bf p}_{3}}{(2 \pi \hbar)^{3}} \int \frac{ d {\bf p}_{4}}{(2 \pi \hbar)^{3}} \nonumber \\
& \times& (2 \pi \hbar)^{3} \delta \left( {\bf p}_{2} - {\bf p}_{3} - {\bf p}_{4} \right) 
%\nonumber \\ & & 
\delta \left( \varepsilon_{c}+\varepsilon_{2}-\varepsilon_{3}-\varepsilon_{4} \right)
\nonumber \\ %& &
&\times& \left[ (1+N_{2})N_{3}N_{4} + N_{2} (1+N_{3})(1+N_{4}) \right].
\label{Sigma_K}
\end{eqnarray}
Although the occupation numbers, $N_{i}$, and therefore also energies $\varepsilon_i$ are in general time-dependent (their values governed by a quantum Boltzmann equation and self-consistent Hartree-Fock energies \cite{Stoof_Langevin,MyTutorial}), current numerical implementations only focus on near-equilibrium situations, for which the thermal cloud is treated as static; in this case, $N_{i}$ can be approximated by the usual Bose-Einstein distributions $N_{i}=\left[\exp\left(\beta \left(\varepsilon_{i}-\mu)\right)\right)-1\right]^{-1}$, where  $\beta = {1}/{k_{B}T}$ and ${\bf p}_{i}$ labels the momentum of a particle in the $i$th single particle energy level. 

As one might expect, the strength of the dissipative effects arising from dynamical particle exchange between the two subsystems is `balanced' by the magnitude of fluctuations present, a relation encapsulated by the fluctuation-dissipation relation for the system, upon noting explicitly the dependence of these quantities on the system energy $\varepsilon_{c}$, namely
%\begin{equation}
$
%\label{fluct_disp_exact}
iR({\bf r};\varepsilon_{c})=-(1/2) \hbar \Sigma^{K}({\bf r};\varepsilon_{c})\left[ 1 +2N(\varepsilon_{c}) \right]^{-1}.
%\end{equation}
$
%While (\ref{fluct_disp_exact}) is exact, as discussed in [DuineStoof,StoofBijlsma2001], 
In order to make the solution of (\ref{Langevin}) numerically tractable, in first instance one can allow the system to equilibrate to a classical distribution by noting that for large occupation numbers, $\left[1+2N(\varepsilon_{c})\right]^{-1}\approx{1}/{2}\beta(\varepsilon_{c}-\mu)$.
%\begin{equation}
%\label{fluct_disp_class}
%\left[1+2N(\varepsilon_{c})\right]^{-1}\approx\frac{1}{2}\beta(\varepsilon_{c}-\mu).
%\end{equation}
Since the condensate energy is still an operator of the form $\varepsilon_{c}=\left( - {\hbar^{2} \nabla^{2} }/{2m} + V_{\rm ext}({\bf r}) + g |\Phi({\bf r};t)|^{2} \right)$,
%\begin{equation}
%\varepsilon_{c}=\Bigg[ - \frac {\hbar^{2} \nabla^{2} }{2m} + V_{\rm ext}(r) - \mu + g |\Phi(r;t)|^{2} \Bigg]
%\end{equation}
the Langevin equation takes the somewhat simplified form
\begin{eqnarray} 
\label{SGPE}
i \hbar \frac{ \partial \Phi({\bf r};t) }{ \partial t} 
=  (1+\frac{\beta}{4}\hbar\Sigma^{K}({\bf r};t)) \Bigg[ - \frac {\hbar^{2} \nabla^{2} }{2m} + V_{\rm ext}({\bf r}) - \mu
\nonumber + g |\Phi({\bf r};t)|^{2} \Bigg] \Phi({\bf r};t) + \eta({\bf r};t)\;.
\end{eqnarray}

%This modified form of the SGPE for the evolution of $\Phi(r;t)$ in contact to a heat bath at fixed temperature has been used extensively to date by Stoof and collaborators (including one of us, NPP). It should be noted that related work has been undertaken by Gardiner, Davis and co-workers [ ].
%%
%%---- Third paragraph ------------------------------------------------------------------------------------
%%

%\noindent 
The very first numerical implementation of this SGPE in the context of ultracold Bose gases was undertaken in 2001 by Stoof and Bijlsma \cite{Stoof_Langevin} to model the reversible growth of a condensate, as seen in the MIT experiment of Stamper-Kurn et al. \cite{MIT_Dimple}: in this experiment, a dimple potential was introduced in a reversible manner to the centre of a harmonic trap, thereby inducing localised phase space compression, and leading to a (periodic) crossing through the phase transition.
Good qualitative agreement was demonstrated between the numerical results \cite{Stoof_Langevin} and the observed lagging behind of the condensate growth, relative to the addition of the dimple trap, found in the experiment. The method, which is also amenable to analytic variational calculations of stochastic dynamics \cite{Stoof_Duine,Stoof_Vortex},
has since been applied (in collaboration with one of us, NPP) to numerical studies of coherence 
\cite{Proukakis_Equilibrium,Low_D_PRA}, atom laser \cite{SGPE_AtomLaser}, and atom chip dynamics \cite{Proukakis_AtomChip}.
It should be noted that related work has been recently undertaken by Gardiner, Davis and co-workers \cite{SGPE_I,SGPE_II,SGPE_III,SGPE_IV,Davis_SGPE_New,Davis_SGPE_Nature}.

The aim of this paper is to briefly review the main concepts and illustrate the versatility of the SGPE formalism for describing systems in which fluctuations are of great importance; for numerical convenience, most of our discussion is restricted to effectively one-dimensional systems, for which we also consider the reduction of the SGPE to simpler theories, which incorporate finite temperature effects, to differing levels of approximation. The extension of this approach to two-dimensional systems is also briefly reviewed. Although, by construction, the results of such simulations are best interpreted after numerical averaging over the different noisy initial conditions, we shall see that important information is also contained in single runs, as also noted in \cite{Davis_SGPE_New,Davis_SGPE_Nature}.

\section{One-Dimensional Applications}
\label{sec:1d}

\subsection{Spatiotemporal Evolution of Coherence in a One-dimensional Bose Gas}

\subsubsection{Non-equilibrium Coherence}

%For a one-dimensional Bose gas, the usual condition for Bose-Einstein condensation, the establishment of off-diagonal long range order in the one-body density matrix[PenroseOnsgager1956], is no longer satisfied. 
Low dimensional systems exhibit a richer phase diagram than their three-dimensional counterparts: in the case of a purely 1d homogeneous system, long wavelength fluctuations in the phase \cite{Popov}, prevent the onset of Bose-Einstein condensation, even down to zero temperature. For a trapped, one-dimensional gas, however, the long-wavelength physics is altered due to the low energy cut-off imposed by the harmonic trapping potential, and the quantum degenerate regime is instead described by two characteristic temperatures, $T_{d}$ and $T_{\phi}$, below which density and phase fluctuations, respectively, become suppressed \cite{Low_D_Petrov}. The intermediate regime \cite{Low_D_PhaseDiagram}, is found to be one in which density fluctuations no longer dominate, yet fluctuations in the phase remain; this is the regime in which the system is said to contain a quasi-condensate \cite{Popov}.

Under tight transverse confinement, the description of the dynamics of highly elongated three-dimensional systems effectively reduces to consideration of a one-dimensional equation, as dynamics in the transverse direction becomes restricted to the lowest mode. As this is the case in many current experiments with atom chips \cite{Low_D_Experiment_2}, it is crucial to fully understand and characterize such fluctuations.  
%In addition, as a key indication for the onset of condensation, an accurate description of fluctuations is essential in describing the effects of temperature on quasi-condensate and coherence growth within low-dimensional systems. Various mean field methods have been devised to this end, with the incorporation of density fluctuations in the low-temperature limit given in [Petrov et el.], and the extension to finite temperatures in Anderson et al[], the latter of which tested favourably against the findings of the stochastic method presented here[NickPRA06]. 
The SGPE is well suited to investigations of phase-fluctuating, or quasi-condensates, as  fluctuations about the mean field are implicitly retained within the order parameter, thus giving access to information on coherence, available through the evaluation of various temporal and spatial numerical auto-correlation functions. 
%This is in complete analogy with experimental techniques [], and forms a powerful technique as a basis for comparison between theory and experiment. In addition, as the order parameter describes the condensate mode plus low lying modes, there is no distinction made between these components which are naturally included within the formalism.

%FIG1
\begin{figure}[t]
  \begin{center}
    \includegraphics[scale=0.55]{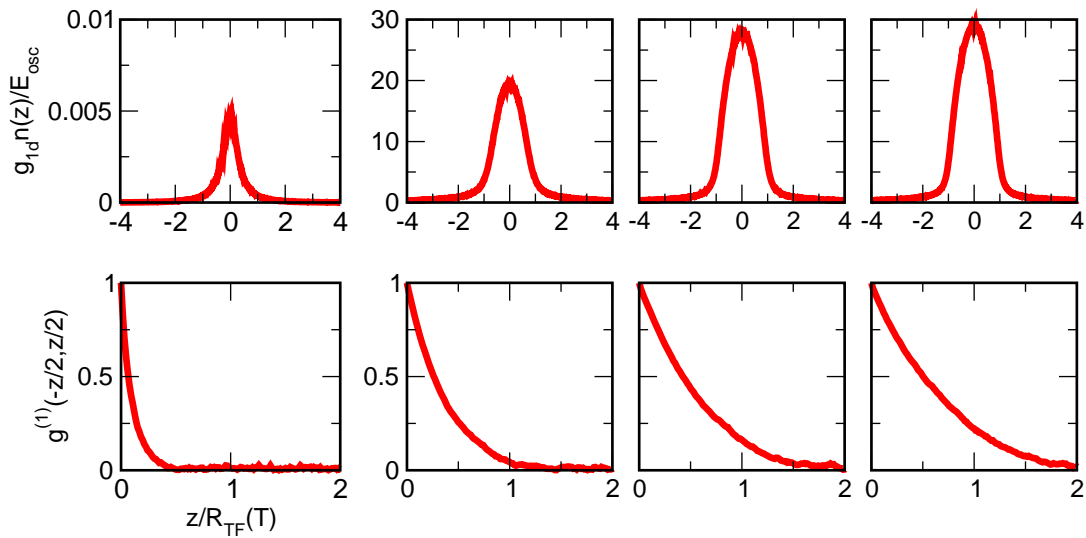}
    \caption{Snapshots of the density and first-order correlation function during the growth of a one-dimensional $^{23}$Na quasi-condensate. The simulations were performed for an effective chemical potential $\mu = 30 E_{\rm osc}$, where $E_{\rm osc} = \hbar \omega_z$ is the harmonic oscillator energy, with $\omega_{z}=2\pi\times30$Hz at a temperature $T=100$nK for a final quasi-condensate atom number $N\approx20000$.
The effective 1D interaction strength $g_{1d}$ is obtained by averaging over transverse Gaussian profiles of width $l_{\perp} = \sqrt{\hbar /m \omega_{\perp}}$, where $\omega_{\perp}=2\pi\times120$Hz. For the chosen somewhat artificial parameters, the phase degeneracy temperature is relatively high, slightly in excess of $100$nK.
Displayed snapshots were taken at $t=0.05t_{eq}$, $t=0.15t_{eq}$, $t=0.3t_{eq}$ and $t=0.45t_{eq}$, where $t_{eq}$ denotes an approximate time for the system to reach dynamical equilibrium.}
    \label{T_100_prof_coh}
  \end{center}
\end{figure}

A study of the lowest order correlation functions of the gas reveals crucial information about the system. In particular, information about the phase of the system is contained in the first order correlation function, with the next higher order correlation containing crucial details about density fluctuations. Such quantities can be directly measured in interferometric `juggling' experiments \cite{Orsay,Hannover1,Hannover2}, and can also be easily obtained numerically within this formalism. More specifically, the renormalised first order correlation function at time $t$ can be obtained in `symmetric form' about a point $z_0$, via the relation
%
%associated with fluctuations in the condensate phase, is defined as follows
\begin{equation}
g^{(1)}(z_0-z/2,z_0+z/2;t)=\frac{\langle\Phi^{*}(z_0-z/2,t)\Phi(z_0+z/2,t)\rangle}{\langle|\Phi(z_0-z/2,t)|^{2}\rangle\langle|\Phi(z_0+z/2,t)|^{2}\rangle} \,.
\end{equation}
%
%Such a quantity can be directly measured in interferometric experiments, such as
% those discussed in \cite{Orsay,Hannover1,Hannover2}. 
%A comparison between this and the asymmetric form $g^{(1)}(z,0;t)$ showed good agreement between the two for high temperatures ($T=2.8 T_{\phi}$) and for spatial points deep within the condensate region ($z<R_{TF}(T)$)[NickPRA06].

We start our discussion by the basic
demonstration of the stochastic method, investigating the nonequilibrium evolution of the density and first-order correlation function during the growth of a one-dimensional quasi-condensate from a static thermal cloud. Fig. \ref{T_100_prof_coh} shows snapshots of the time evolution of both the density (top row) and coherence (bottom row) towards their equilibrium values, for a gas in contact with a heat bath at a fixed temperature.
% $T\approx0.9T_{\phi}$. We see that 
%As the system heads towards equilibrium, the spatial extent of the correlation function increases, until it converges to some finite value.
% across the whole system size for later times.
As the atomic system is cooled and heads towards the new equilibrium, the spatial extent of the correlation function increases as phase coherence is established, with the final shape attained dictated by the temperature, for a given particle species, number and trap geometry. 
In this and all subsequent figures on coherence, distances are scaled to the effective quasi-condensate spatial extent $R_{TF}(T)$ at equilibrium. This quantity takes account of the varying system size at different temperatures due to quasi-condensate depletion, and reduces at $T=0$ to the `usual' Thomas-Fermi solution $R_{TF}(T=0)=\sqrt{2 \mu/m \omega_z^2}$. There are two ways to obtain $R_{TF}(T)$, as discussed in \cite{Proukakis_Equilibrium}: firstly, by considering the modified low-dimensional theory of Andersen {\em et al.} \cite{Low_D_Stoof}, or equivalently by using the curvature of the density profiles to obtain a corresponding quasi-condensate size.
The particular example considered here corresponds to the case where the coherence length of the system is comparable to the system size, with $T\approx0.9T_{\phi}$.
%It is interesting to note, as pointed out in [StoofBijlsma2001], that without the noise as a seed, no growth would take place at all and the density would be trivially zero.
The observed growth in the densities is a direct consequence of the transfer of particles from the heat bath to the system, and the noise is crucial in order to seed the growth process \cite{Stoof_JLTP,Stoof_Langevin}.

%\subsubsection{Equilibrium Coherence}

For temperatures such that $T>T_{\phi}$, the first-order correlation function has been found experimentally to exhibit exponential decay, whereas for $T<T_{\phi}$ the corresponding spatial dependence becomes Gaussian \cite{Orsay,Hannover1,Hannover2}. 
%As the atomic system is cooled and heads towards the new equilibrium, the spatial extent of the correlation function increases as a degree phase coherence is established, with the final shape attained dictated by the temperature at this equilibrium, for a given particle number and trap parameters. 
%
Thus, one possible way to quantify the degree of coherence and extract a uniformly-varying coherence length is based on a simple fitting function of the form \cite{Proukakis_Equilibrium}, $f(z)=\exp\{-[\left({z}/{L_{coh}}\right)+\zeta\left({z}/{L_{coh}}\right)^{2}]\}$,
%\begin{equation}
%\label{fit}
%f(z)=\exp\left\{-\left[\left(\frac{z}{L_{coh}}\right)+\zeta\left(\frac{z}{L_{coh}}\right)^{2}\right]\right\}
%\end{equation}
in which $\zeta$ is a parameter characterizing the crossover from exponential to Gaussian behaviour.
% with decreasing temperature. From this fit, we can extract the quantity $L_{coh}$, which gives a universal coherence length for the system within each regime.

\begin{figure}[t]
  \begin{center}
    \includegraphics[scale=0.4]{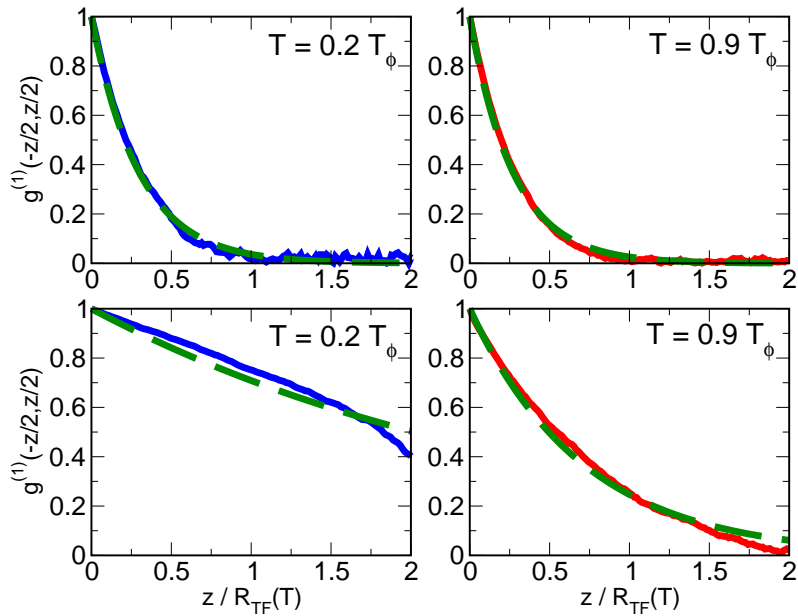}
    \caption{Coherence growth at the trap centre for the temperatures $T=0.2T_{\phi}$ and $T=0.9T_{\phi}$. The upper row shows the coherence evaluated at a time early in the growth, $t=0.15t_{eq}$, for each temperature, whereas the plots in the lower row display the coherence at the equilibrium time, $t=t_{eq}$.
Solid (noisy) curves correspond to numerical results obtained after suitable averaging over a few thousand runs, whereas dashed lines indicate the `optimum' fit using the function $f(z)$ given in the text. Other parameters are as in Fig.\ 1.
}
    \label{coh_growth_profs}
  \end{center}
\end{figure}

A comparison between the spatial coherence at the trap centre (i.e.\ for $z_0 =0$) for early and late times, together with the results of the fitting procedure, is shown in Fig. \ref{coh_growth_profs} for two temperatures, corresponding roughly to the case of a 'pure' condensate (left images)  and a gas at the crossover from `pure' to `quasi'-condensation (right images).
%The spatial system size is scaled to a temperature dependent radius, $R_{TF}(T)$, which was determined from one-dimensional mean-field theory [].

%Fig2

As is evident from this figure (right images), for 
 the higher temperature results, corresponding to the parameters of Fig.\ \ref{T_100_prof_coh}, the equilibrium correlation function decays more rapidly than for the lower temperature limit.
%, indicating phase fluctuations are not sufficiently suppressed for a universal phase to be established. For the lower temperature, phase fluctuations are well suppressed and the system can be said to have undergone 'true condensation', defined in the sense that the coherence stretches now well beyond the system size. 
%
Although the spatial extent of the system differs between these two cases ($R_{TF}(0.2T_{\phi})= 7.4 l_z$ vs.\ $R_{TF}(0.9T_{\phi})= 6.6 l_z$), the appropriately scaled correlation function plotted in the top row of Fig.\ \ref{coh_growth_profs} exhibit very similar behaviour, unlike the case at their respective equilibria (bottom row).
Nonetheless, the chosen fitting function seems to appropriately capture the main features; importantly, this enables us to study in a universal manner the growth of coherence as the system is equilibrating, being driven by the heat bath.

Numerous experiments \cite{MIT_Formation,QK_PRL_III,Growth_Amsterdam,Growth_Zurich} have been performed to study the relation between condensate growth and evolution of coherence, and a detailed comparison to such experiments is pending 
(but see also \cite{QK_JPhysB,Growth_Orsay} and references therein). For our present illustrative purposes, we simply highlight that the growth of coherence appears to exhibit similar behaviour to typical quasi-condensate number growth curves, although these generally follow such curves with an additional small lagging time; most notably, once the norm has reached its steady-state value, the coherence is still growing, albeit at a very slow rate.

\begin{figure}[h]
  \begin{center}
%    \begin{tabular}{cc}
      \includegraphics[scale=0.4]{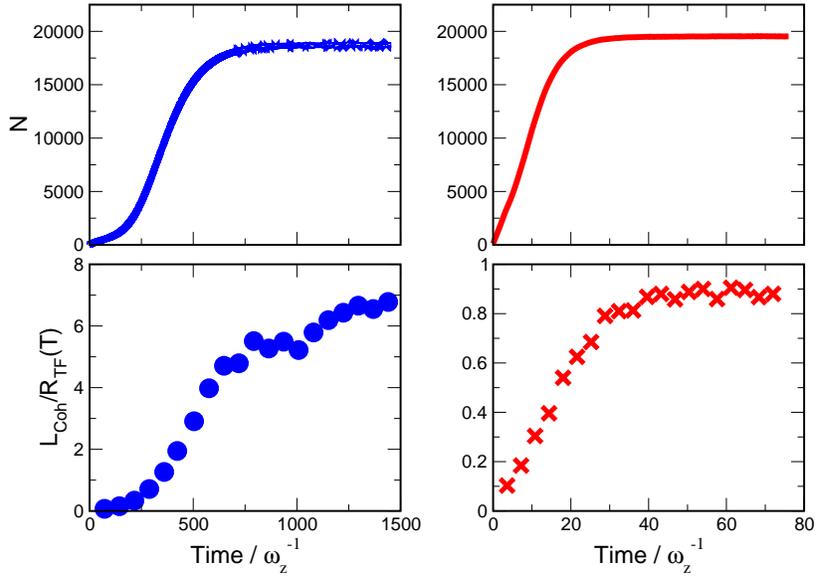}
%      \resizebox{65mm}{!}{\includegraphics{}} &
%      \resizebox{65mm}{!}{\includegraphics{}} 
%    \end{tabular}
    \caption{
Quasi-condensate atom number growth curves for a system equilibrating in contact with a prescribed heat bath (top row), versus corresponding temporal evolution of the coherence length $L_{coh}$, obtained to approximately $10\%$ accuracy from the fitting function defined in the text (bottom row). Note the differing y-axis scales between the coherence growth for two temperatures, with results scaled to the temperature dependent Thomas-Fermi radius $R_{TF}(T)$.}
    \label{coh_growth}
  \end{center}
\end{figure}

With this means of quantifying the degree of coherence across a sample, it is then interesting to assess how $L_{coh}$ varies with time during the cooling process.
The analysis of these results for the two different temperatures in the context of this somewhat idealised scenario is plotted in Fig. \ref{coh_growth}.
As these simulations are based on a fixed value of $\mu$, the final atom numbers are slightly increased in the high temperature case (top right image); note also the significantly increased relaxation time and observation of the initial spontaneous initiation process prior to the domination of the bosonic stimulation effect (top left).

%FIG 3
%FIG 4
\begin{figure}[hb!]
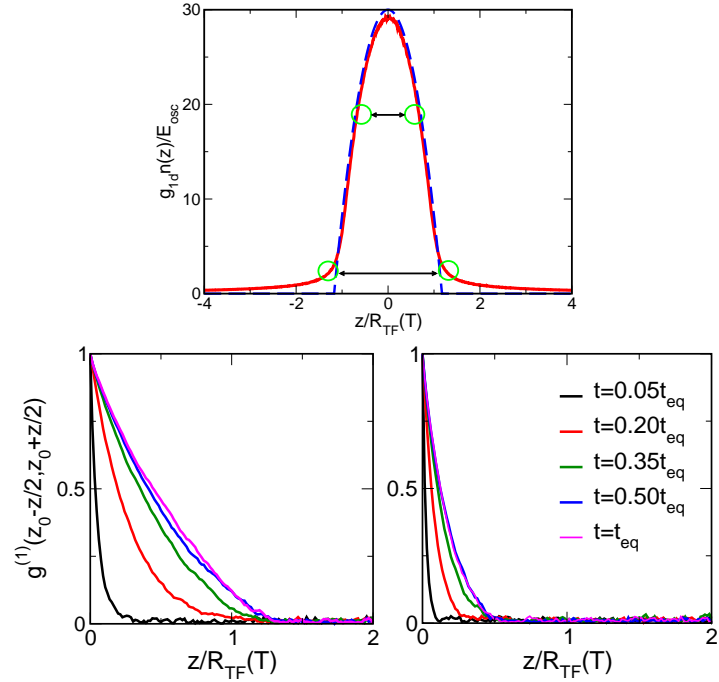

  \begin{center}
    \begin{tabular}{c}
      \includegraphics[scale=0.25]{T_100_profile.eps} \\
      \includegraphics[scale=0.35]{T_100_g1_LeftColumnData2.eps} 
    \end{tabular}
    \caption{Evolution of the coherence (from left to right within each image) during the growth to equilibrium with the heat bath for a temperature $T=0.9T_{\phi}$. The correlation function is evaluated at $z_{0}=0.7R_{TF}(T)$ (lower left plot) and $z_{0}=1.4R_{TF}(T)$ (lower right plot), with $z_0$ indicated by the circles on the equilibrium density profile for the chosen temperature (top plot).}
    \label{coh_position}
  \end{center}
\end{figure}

\subsubsection{Dependence of Equilibrium Coherence on Position}

We also investigate the position dependence of the correlation function at equilibrium, by comparing the evolution of $g^{(1)}$ at different positions $z_{0}$, for a fixed temperature corresponding to the high temperature case presented in the above results. The results of this investigation are shown in Fig. \ref{coh_position} for $z=0.7R_{TF}$ and $z=1.4R_{TF}$  for various times during the growth to equilibrium.
The final profile attained at this temperature, together with the Thomas-Fermi profile for the same number of atoms, is also shown, with the positions considered indicated.
The coherence within the Thomas-Fermi radius decreases uniformly but at a relatively slow rate as soon as one probes off-centred regions, with the effective coherence length decreasing rapidly as soon as one moves beyond the Thomas-Fermi radius (right images).

\subsubsection{Potential 'Identification' of a Quasi-condensate}

\begin{figure}[b]
  \begin{center}
%    \begin{tabular}{cc}
      \includegraphics[scale=0.4]{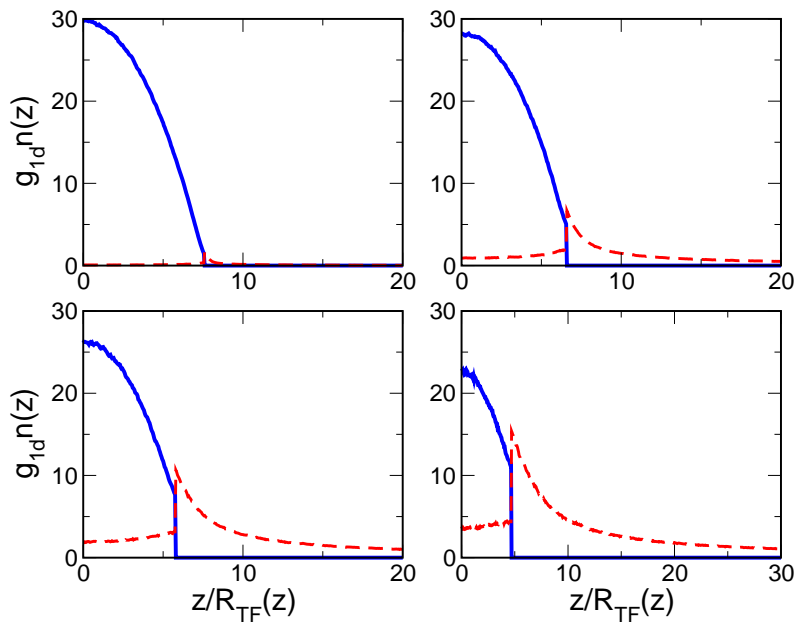}
%      \resizebox{65mm}{!}{\includegraphics{}} &
%      \resizebox{65mm}{!}{\includegraphics{}} 
%    \end{tabular}
    \caption{Quasi-condensate (solid) and thermal (dashed) density profiles obtained from the SGPE by manipulation of density-density correlations. Plotted images correspond to the following temperatures $T/T_{\phi}=$ 0.1, 0.9, 1.8, and 2.9 (top left to bottom right).}
    \label{split_prof}
  \end{center}
\end{figure}

An appealing feature of the stochastic approach that is often not fully appreciated, is that it actually generates total density profiles without needing to resort to a special mean-field treatment of the condensate mode. This means that, as in the experiments, one does not have an automatic 'visualisation' of the condensate in the system. So, in order to analyse these results, one may consider applying bimodal fits precisely as done in the experiments. However, an alternative definition has been proposed \cite{Svistunov_QC}, in which the quasi-condensate density is identified as 
%\begin{equation}
$
n_{\textrm{QC}}(z)=\sqrt{\ave{|\Phi(z)|^{2}}^{2}-\ave{|\Phi(z)|^{4}}}.
%\label{quasi_split}
%\end{equation}
$
This approach has been implemented in \cite{Proukakis_Equilibrium},
with the density profile contribution for the thermal cloud over low-lying modes determined self-consistently via the relation
$
%\begin{equation}
%\label{therm_split}
n_\textrm{T}(z)=\ave{|\Phi(z)|^{2}}-n_{\textrm{QC}}(z),
%\end{equation}
$
within the region $z<R_{TF}(T)$, and given simply by $n_T(z)\equiv n(z)$ outside of the Thomas-Fermi radius.
%which there is only a thermal contribution, so $n(z)\equiv n_{T}(z)$.
%
The identification of these quantities enables spatial plots of the temperature dependent quasi-condensate density profiles to be obtained in an \emph{ab initio} manner, as shown for different temperatures in Fig. \ref{split_prof}.
Such a definition was used more recently to distinguish between a condensate and a quasi-condensate \cite{Davis_SGPE_New,Bisset_Rot_cond}

%FIG 5

Having discussed some basic elements of finite temperature trapped, quasi-one-dimensional Bose gases via the SGPE, we now discuss certain limiting cases of the above theory which are commonly used in the literature, and which
shed more light on the benefits and versatility of the SGPE.
%move on to possible simplifications, which lead to contact with other theories.

\subsection{Reduction to Alternative Theories}

%The finite temperature dynamics of trapped Bose gases are well represented within the SGPE formalism which can also be shown to reduce to a number of sub-theories, each capturing different physics pertaining to finite temperature effects. We describe in the following sections, firstly the norm conserving evolution of noisy initial conditions, corresponding to the removal of contact with the heat bath, and secondly the dissipative GPE, in which the self energy is found to provide a microscopically justified damping at a given temperature.

We have already argued that the two important features of the SGPE are that it includes the correct damping due to its dynamical coupling to the heat bath, and that it extends beyond mean field effects, by including a stochastic noise term. Removing either (or both) of these elements leads to simpler theories \cite{MyTutorial}, which can nonetheless be useful in certain regimes, as discussed below.

\subsubsection{Removing Contact with the Heat Bath}

In first instance, we can consider what happens when the coupling to the heat bath is abruptly (and perhaps somewhat artificially) removed. We recall that
upon beginning a simulation with the SGPE, the effect of the $iR(r;t)$ term in \eqref{Langevin} is to pump particles from the heat bath, and populate the low lying modes to their equilibrium occupations, in accordance with the classical equilibrium for a given set of trap frequencies, chemical potential and temperature. On reaching a dynamical equilibrium, the scattering rate $\propto iR(r;t)$ becomes on average equal to zero, and there is no net particle exchange with the heat bath, although this continuous exchange can cause additional damping. However, any externally imposed perturbations of the system about this set of system parameters, would typically lead to the introduction or loss of particles, upon further evolution with \eqref{Langevin}.

In general, the SGPE is coupled to a quantum Boltzmann equation for the population of the higher-lying modes, and these two equations should be solved self-consistently. If this is not done, then modifying for example the trap geometry will lead to a change in the norm of the SGPE. If we are only considering the evolution of the low-lying modes, in order to keep the atom number within our system fixed, one could consider `fixing' the norm in the numerics by removing the coupling to the heat bath. A better alternative might perhaps be to change the effective chemical potential in the SGPE to the final value corresponding to the same atom number in the new trap, although the observed dynamics will typically depend on how quickly the chemical potential is adjusted to its final value.

%In the former case, the numerical procedure amounts to propagating the stochastic equation to its equilibrium solution, before suddenly switching off both the noise and the damping terms of \eqref{Langevin} can be removed to maintain explicit norm conservation. 
One procedure is thus to propagate initially the full stochastic solution in order to obtain the desired equilibrium state, before switching off the noise and damping terms of \eqref{Langevin}, and subsequently propagating the noisy initial condition via the usual GPE which enforces particle number conservation. This corresponds physically to the removal of contact with the heat bath upon reaching equilibrium.

% and so particle number conservation in the subsequent dynamics. This procedure still must be repeated, however, for multiple initial field configurations in order to accurately sample the phase space probability distribution.

As a method, this procedure has strong parallels with the Truncated Wigner method, and a comparative study of the two is under way currently \cite{SGPE_vs_TWA}. 
The main difference arises in the fact that in the usual truncated Wigner implementations (see, e.g. \cite{Steel_TWA,TWA_Norrie_PRL,Scott_Hutchinson_Gardiner} or \cite{MyTutorial} for a more extensive list), only quantum (as opposed to thermal) noise is typically added in the initial conditions (but see also \cite{Castin_TWA,TWA_Ruostekoski_2}), whereas the SGPE in general contains both quantum and thermal fluctuations \cite{Stoof_JLTP}. Although existing numerical implementations of the SGPE focus on the thermal effects (due to the `classical' approximation mentioned earlier), the important advantage of this approach is that
%
%As mentioned, within the truncated Wigner approach, quantum fluctuations are approximately introduced via the initial conditions, through appropriate sampling of the Wigner distribution. This amounts to the introduction of an extra half a particle per mode, on average, to the matter field, whose dynamical evolution is then dictated by the GPE; there is a formal equivalence between the propagation of this field by the GPE and the evolution of the Wigner function, under a suitably truncated equation of motion []. As the GPE is a classical field equation, however, the system will equilibrate, through the nonlinear mixing of modes, from the initial quantum distribution a particular temperature, to a higher temperature classical distribution. Through 
there is an explicit fluctuation-dissipation relation which guarantees convergence to the correct classical equilibrium at any pre-determined temperature; in other words, the initial thermal state for the unperturbed system can be correctly assigned.
The validity of conventional truncated Wigner implementations is on the other hand limited to very low temperatures \cite{Castin_TWA} and occupation numbers for which quantum fluctuations have a dominant role - see \cite{SGPE_vs_TWA} for a more detailed discussion.
%FIG6
\begin{figure}[h!]
  \begin{center}
%    \begin{tabular}{cc}
      \includegraphics[scale=0.5]{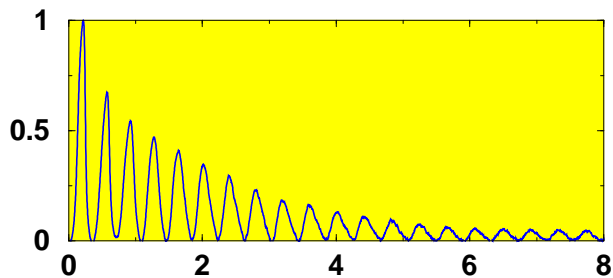}
%      \resizebox{65mm}{!}{\includegraphics{}} &
%      \resizebox{65mm}{!}{\includegraphics{}} 
%    \end{tabular}
    \caption{Suppression of shockwaves induced upon the introduction of a deep dimple trap to a one-dimensional ultracold Bose gas of around 3500 $^{23}$Na atoms, as manifested in the temporal evolution (in units of the inverse trap frequency) of the corresponding damped central density oscillations obtained after numerical averaging, and scaled to the peak density for the given system. (Details of other parameters used in these simulations can be found in \cite{Proukakis_AtomChip}.) Contact with the heat bath was removed prior to the introduction of the dimple trap to maintain number conservation.}
    \label{shock_damp}
  \end{center}
\end{figure}

This finite temperature approach based on the removal of the heat bath once the correct equilibrium has been obtained via the SGPE
% This technique 
was used to probe the dynamics of an initially equilibrated thermal stationary state following the addition of a deep Gaussian dimple trap \cite{Proukakis_AtomChip}. 
A large and sudden perturbation is found to lead to the appearance of shock waves in the system, and in the averaged profiles typically studied, these manifest themselves in terms of an initial increase in the central density, followed by periodic damped decrease-increase cycles, as shown in Fig.\ref{shock_damp}; this can be interpreted as direct evidence of the temperature-induced suppression of the initial shockwave amplitude induced by the addition of the new trap.
It should be noted here that the introduction of the correct thermal noise in the initial conditions is crucial to the generation of the damping which would not be present in simulations of the ordinary GPE starting from a stationary condensed state.

\subsubsection{A Microscopically-based Dissipative GPE}

An even simpler limiting case of the SGPE is based on
removing only the noise term of \eqref{Langevin}; then, the SGPE reduces to a `dissipative GPE', as noted for example in \cite{SGPE_III,QK_PRL_IV}. The form of the damping is given by the self energy, and as such is spatially dependent, consistent with the spatially inhomogeneous background of the harmonic trap. 
Such a model has been used to study vortex and soliton dynamics leading to a number of interesting predictions. However, the origin of such models can be traced to rather crude approximations in the more advanced microscopic models, such as the SGPE, with most such publications justifying this equation phenomenologically, and actually using a constant value for the damping $\gamma$ \cite{Choi_Phenomenology,Vortex_Lattice_Ueda_1,Vortex_Lattice_Ueda_2,Proukakis_Parametric}.

%FIG 7
\begin{figure}[htp]
  \begin{center}
%    \begin{tabular}{cc}
      \includegraphics[scale=0.85]{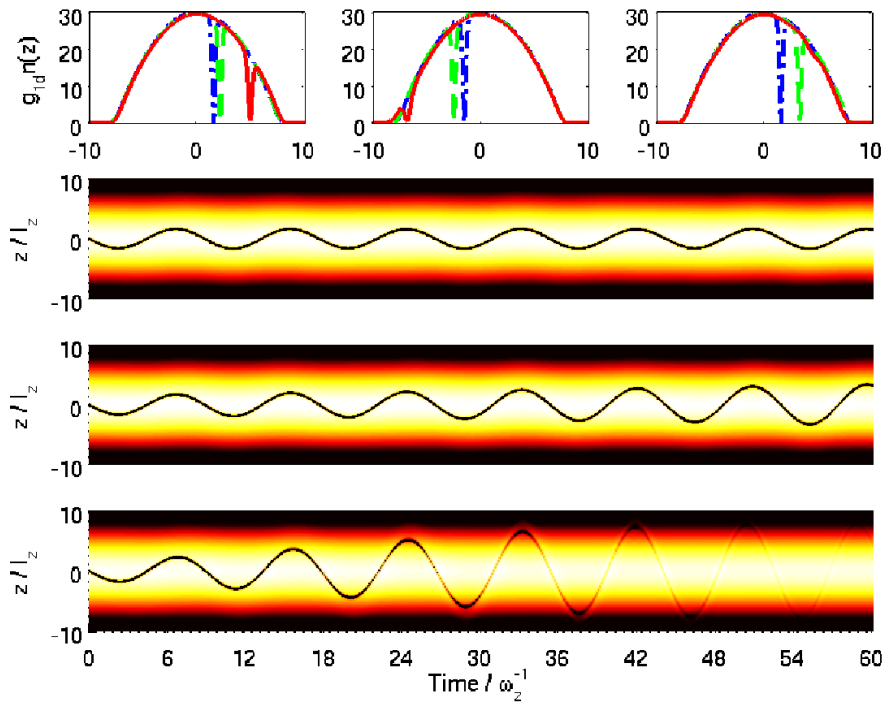}
%      \resizebox{65mm}{!}{\includegraphics{}} &
%      \resizebox{65mm}{!}{\includegraphics{}} 
%    \end{tabular}
    \caption{Soliton dynamics within a one-dimensional gas of $N\approx 20000$ $^{23}$Na atoms, for the zero temperature (GPE) case, plus at temperatures of 100nK and 200nK. The top row plots show density snapshots at three times during the dynamics, $t=24\omega_{z}^{-1}$, $t=37.2\omega_{z}^{-1}$ and, $t=60\omega_{z}^{-1}$; shown in each figure are the results for all three temperature cases at that time (zero temperature - blue, dot-dashed line; 100nK - green, dashed line; 200nK - red, solid line). The spacetime plots are ordered downwards in order of increasing temperature, for the highest temperature case the soliton can be seen to decay rather rapidly.}
    \label{sol_damping}
  \end{center}
\end{figure}

The dissipative equation which results from the SGPE  in general takes, within the `classical approximation', the form
\begin{eqnarray} 
i \hbar \frac{ \partial \Phi({\bf r};t) }{ \partial t} 
=  (1-i\gamma({\bf r},T,t)) \Bigg[ - \frac {\hbar^{2} \nabla^{2} }{2m} + V_{\rm ext}({\bf r}) - \mu
+ g |\Phi({\bf r};t)|^{2} \Bigg] \Phi({\bf r};t)\;,
\label{disp_GPE}
\end{eqnarray}
where the damping term (for a system made up of a fixed total number of atoms) is given by
\begin{equation}
\gamma({\bf r},T,t)=i\frac{\beta}{4}\hbar\Sigma^{K}({\bf r},t) \,,
\end{equation}
although our discussion so far has additionally ignored the time-dependence for numerical simplicity.

It is very easy to see that
the resulting evolution is no longer norm conserving, and it therefore becomes nessessary to renormalize the wavefunction as each time step to maintain a constant particle number. 

For illustrative purposes, we apply this approach here to model the oscillations of an `ideally generated' dark soliton within a one-dimensional condensate. 
The dynamics for three identical initial soliton solutions are shown at different temperature in Fig. \ref{sol_damping} (top images), with each image showing the position of the dark soliton for three different temperatures, corresponding to a different evolution time. 
%The initial condition for each case is the same, with the soliton generated with speed $v=0.2c$, where $c$ is the speed of sound within the condensate, although the subsequent evolution is subject to different damping rates due to the different temperatures considered here. 
To best illustrate the different temperature-dependent decay rates, we give corresponding
%
%The dynamics for three identical initial soliton solutions are shown different temperature in Fig. \ref{sol_damping} where the decay is illustrated by a continuous 
space-time plots of the density along the z-axis
%together with density snapshots taken during the evolution for each temperature. Behaviour of this type has been 
with such damping already observed in experiments \cite{Hannover_Soliton_Exp}.
% and it was shown that the damping found could be explained on the basis of finite temperature effects in [BrianNickPRA2007].

Although consideration of the 1D case enables us to bring out the main physical features of the SGPE model, comparison to experiments requires us to consider its properties in higher-dimensional systems, and we conclude our presentation by a simple illustration in two-dimensional geometries, and some related comments.

%\newpage
\section{Extension to Two Dimensions}
\label{sec:2d}

%\noindent In this section, we discuss the spontaneous appearance of topological defects such as vortices in the context of SGPE simulations, which are more easily discernible than in one-dimensional simulations. In addition, we will mention briefly some of numerical issues associated with the application of the formalism in higher dimensions, investigated for the two-dimensional case.

As in one dimension, for a uniform two dimensional Bose gas, thermal fluctuations remain even at very low temperatures, preventing the onset of Bose-Einstein condensation 
\cite{Mermin,Hohenberg}. For interacting trapped two dimensional gases, however, quasi-condensation is possible for low temperatures \cite{Petrov2d}, with a Berezinskii-Kosterlitz-Thouless phase transition to a superfluid state associated with the the pairing of vortices of opposite circulation beneath some critical temperature. Recently, this has spawned much interest in the physics of two-dimensional trapped ultracold systems, particularly in assessing the nature of the phase transition at the critical point \cite{Low_D_Experiment_3,DalibardNJPhys,CornellPRL1,PGPE_BKT_PRL}. 

\subsection{Spontaneous Processes - Vortex Nucleation}

Phase transitions are strongly associated with such topological defects, the appearance of which are predicted by the Kibble-Zurek mechanism 
\cite{Spontaneous_Vortex_Kibble,Spontaneous_Vortex_Zurek}.
An appealing feature of the stochastic approach is that in including fluctuations about the mean field, excitations in the form of spontaneous vortices can be seen during the growth process, as investigated recently using an alternative formulation of the SGPE \cite{Davis_SGPE_Nature}. 
Our discussion so far has been restricted to analysing predictions of the SGPE based on numerical averaging over different initial noisy conditions. However, the monitoring of the evolution within a single run can also provide important information on the system parameters, which may otherwise be lost through averaging - e.g.\ information related to spontaneous events, such as topological defect formation \cite{Spontaneous_Vortex_BEC}.

Within the numerical SGPE applied here, condensate formation from the static thermal cloud is initiated for a positive value of the chemical potential in \eqref{SGPE}. A rapid quench can be simulated by an instantaneous change in the chemical potential, leading to a rapid condensate growth. Within simulations, the appearance of a (random) number of vortices is common, although their details vary from run to run. The appearance of these actually becomes `washed out' following the averaging process, as, due to the random nature of the nucleation of vortices, the positions at which they appear varies between different realisations.
Such studies should be contrasted to studies based on the GPE, in which a vortex is actually `artificially' imprinted at a predetermined position; when modelling the ensuing dynamics, one often introduces damping by resorting to the dissipative GPE $\eqref{disp_GPE}$, with the damping term, $\gamma(x,y)$, which is actually proportional to the two-dimensional self-energy at a particular temperature, typically treated as a constant.
%FIG 8
\begin{figure}[htp]
  \begin{center}
%    \begin{tabular}{cc}
      \includegraphics[scale=0.8]{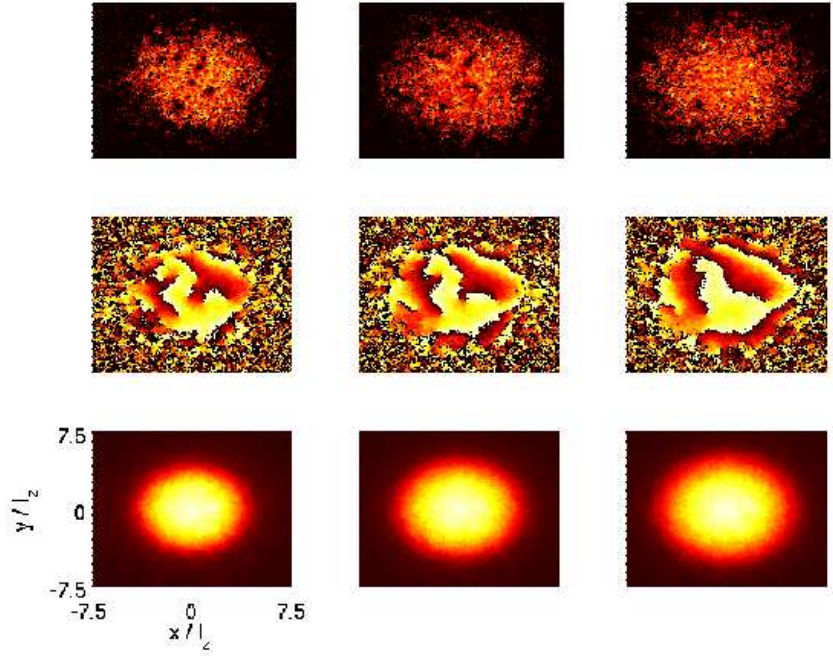}
%      \resizebox{65mm}{!}{\includegraphics{}} &
%      \resizebox{65mm}{!}{\includegraphics{}} 
%    \end{tabular}
    \caption{Nonequilibrium density and phase snapshots for the growth of a two dimensional harmonically trapped quasi-condensate in the x-y plane, for $N\approx60000$ $^{23}$Na atoms, dimensionless coupling constant $g_{2d}=\sqrt{8\pi}a_{3d}/l_{z}\approx0.042$ and trap frequencies $\omega_{x}=\omega_{y}=2\pi\times200$Hz and $\omega_z=2\pi\times4$kHz. The single run density profiles (top row) show the spontaneous appearance of vortices, as apparent also in the single run phase plots (middle row). Time increases to the right in the plots; at the centre of the single run images two vortices can be seen to come together and annihilate. The bottom row profiles are density profiles obtained following averaging over $100$ single runs, with snap-shots taken at the same times as for the single runs. The results show a far smoother density profile is obtained upon averaging, though vortices are now no longer visable.}
    \label{spont_vort}
  \end{center}
\end{figure}

In density plots, vortices appear as dark regions, corresponding to positions at which singularities in the phase of the system can be seen. An example of the results from a typical simulation based on the SGPE is shown in Fig. \ref{spont_vort}. In the density and phase snapshots shown, two spontaneously generated vortices (top left) can be seen to come together (middle) and then annihilate (right), illustrating the spontaneous dynamics captured within simulations. The averaged density profiles included show the result of averaging over $100$ runs, and the smooth density plots which result.

\subsection{Numerical Issues - Dependence on Momentum Cutoff}

%FIG 9
\begin{figure}[htp]
  \begin{center}
%    \begin{tabular}{cc}
      \includegraphics[scale=0.78]{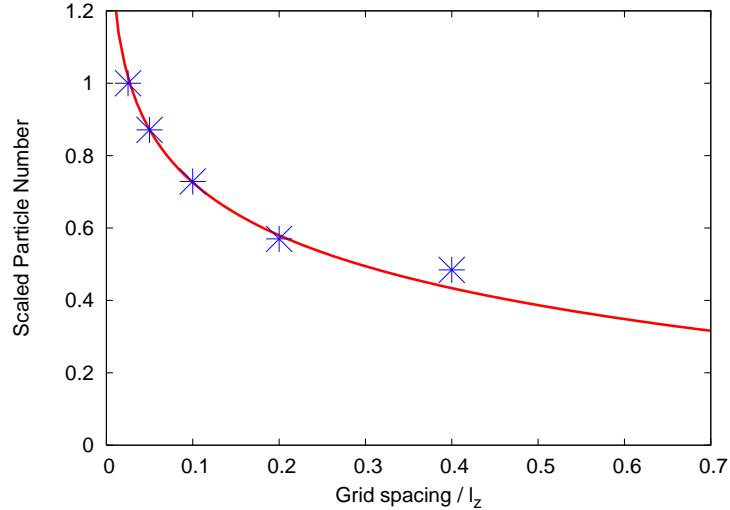}
%      \resizebox{65mm}{!}{\includegraphics{}} &
%      \resizebox{65mm}{!}{\includegraphics{}} 
%    \end{tabular}
    \caption{Dependence of the norm on the numerical grid spacing for the two-dimensional SGPE. The fit shows the dependence of the particle number on the high momentum cut-off, $\Lambda$, introduced by the lattice for fixed chemical potential, temperature and trap frequencies. This is inversely related to the grid spacing $\Delta z$, against which the results are plotted. The divergence is found to be $\propto\log(\Lambda)$ in the two dimensional case, as shown by the line indicating the fit to a function $\propto\log(\pi/\Delta z)$.} 
    \label{2d_div}
  \end{center}
\end{figure}

The numerical means of evolving the system discussed so far make use of the
macroscopic occupation of the low lying modes, assuming these states to be well
described by a classical field. In two dimensions, known divergences arise,
associated with large momenta, or equivalently small grid spacings, as the
continuum limit is approached on the lattice. This is due to the high momentum cutoff, $\Lambda$, related to the numerical grid spacing, $\Delta z$, through the relation $\Lambda=\pi/\Delta z$. 

We present here preliminary results of an investigation into this dependence for the SGPE as a `caution' to the reader. Here we make no adjustments to parameters to compensate for grid dependence and find the expected logarithmic divergence in the momentum cut-off, with the system norm (or number of particles in the classical region) increasing with decreasing spatial discretization $\Delta z$, as shown in Fig. \ref{2d_div}.

In principle such divergences can be removed from the problem through the appropriate renormalization of physical quantities from their original (`bare') values.  For example in homogeneous systems, the lattice effects on temperature or density can be taken into account in an \emph{a posteriori} manner, as described in \cite{Prokov'ev_renorm}. An alternative method is to introduce counter-terms to the system Hamiltonian, which has been successfully applied to Langevin equations previously in, for example, \cite{GleiserPRE2000,GleiserNucB,WojtasThesis,HabibBettencourt}. 

%%
%%---- Conclusions ---------------------------------------------------------------------------------------
%%

\section{Conclusions}

The stochastic Gross-Pitaevskii is emerging as a key approach for the equilibrium and dynamical properties of degenerate Bose gases, and is particularly beneficial for describing regimes of large phase and density fluctuations, such as near (or at) the phase transition, or for weakly-interacting low-dimensional Bose gases. Single run results contain important (at least qualitative) information on spontaneous processes, such as defect formation characteristics, whereas averaged profiles are most suitable for determination of correlation functions, quasi-condensate fractions and smooth density profiles. Although these stochastic simulations have obvious advantages, their existing numerical implementation have not yet explored such theories to their full potential, with current models typically assuming a static (near-equilibrium) thermal cloud, and additionally discarding quantum fluctuations in simulations. Finally, the important issue of the cut-off dependence, already highlighted by other researchers should not be overlooked, particularly for dimensions higher than one, and various techniques are available for renormalising the observed profiles to their actual values, thereby eliminating (at least to leading order) such dependences. We believe that use of the stochastic Gross-Pitaevskii equation, which is also coupled to a quantum Boltzmann equation for the self-consistent dynamical treatment of the higher-lying thermal modes, should be able to describe essentially all main features of weakly-interacting Bose gases currently pursued experimentally, and therefore look forward to the exciting times which lie ahead.

%%
%%---- Acknowledgments -----------------------------------------------------------------------------------
%%

\section*{Acknowledgments}

We are grateful to Henk Stoof for early collaboration and discussion on these ideas and to the UK EPSRC for funding.

%%
%%---- Bibliography --------------------------------------------------------------------------------------
%%

\end{document}